\documentclass[a4paper,twocolumn,10pt,accepted=2022-04-21]{quantumarticle}
\pdfoutput=1
\usepackage[utf8]{inputenc}
\usepackage[english]{babel}
\usepackage[T1]{fontenc}
\usepackage{amsmath}
\usepackage{hyperref}

\usepackage{tikz}
\usepackage{lipsum}
\usepackage[numbers]{natbib}

\usepackage[centering,hmargin=2cm,vmargin=2.5cm]{geometry}

\usepackage{dcolumn}	
\usepackage{bm}			

\usepackage{graphicx}	
\graphicspath{{figures/}{./}} 

\usepackage{braket}
\usepackage{listings}
\usepackage{color}
\usepackage{realboxes}
\usepackage{enumitem}   
\setlist{nolistsep}

\usepackage{float}
\makeatletter
\let\newfloat\newfloat@ltx
\makeatother
\usepackage{algorithm}
\usepackage{algpseudocode}
\algnewcommand\algorithmicswitch{\textbf{switch}}
\algnewcommand\algorithmiccase{\textbf{case}}
\algnewcommand\algorithmicassert{\texttt{assert}}
\algnewcommand\Assert[1]{\State \algorithmicassert(#1)}%

\definecolor{codegreen}{rgb}{0,0.6,0}
\definecolor{codegray}{rgb}{0.5,0.5,0.5}
\definecolor{codepurple}{rgb}{0.58,0,0.82}
\definecolor{backcolor}{rgb}{0.9,0.9,0.88}
\definecolor{mygray}{rgb}{0.8,0.8,0.8}

\lstdefinestyle{mystyle}{
    language=C++,
    backgroundcolor=\color{backcolor},   
    commentstyle=\color{codegreen},
    keywordstyle=\color{magenta},
    numberstyle=\tiny\color{codegray},
    stringstyle=\color{codepurple},
    basicstyle=\small,
    breakatwhitespace=false,         
    breaklines=true,                 
    captionpos=b,                    
    keepspaces=true,                 
    numbers=left,                    
    numbersep=5pt,                  
    showspaces=false,                
    showstringspaces=false,
    showtabs=false,                  
    tabsize=2,
    columns=fixed
}
\def\snippet#1{\Colorbox{backcolor}{\lstinline|#1|}}

\begin{document}

\title{Fast simulation of quantum algorithms using circuit optimization}

\author{Gian Giacomo Guerreschi}
 \email{gian.giacomo.guerreschi@intel.com}
 \affiliation{Intel Labs, Intel Corporation, 2200 Mission College Blvd, Santa Clara, CA 95054}
 \orcid{0000-0002-5579-451X}

\maketitle


\begin{abstract}
Classical simulators play a major role in the development and benchmark of quantum algorithms and practically any software framework for quantum computation provides the option of running the algorithms on simulators.
However, the development of quantum simulators was substantially separated from the rest of the software frameworks which, instead, focus on usability and compilation.
%
Here, we demonstrate the advantage of co-developing and integrating simulators and compilers by proposing a specialized compiler pass to reduce the simulation time for arbitrary circuits.
While the concept is broadly applicable, we present a concrete implementation based on the Intel Quantum Simulator, a high-performance distributed simulator. As part of this work, we extend its implementation with additional functionalities related to the representation of quantum states. The communication overhead is reduced by changing the order in which state amplitudes are stored in the distributed memory, a concept analogous to the distinction between local and global qubits for distributed Schr\"odinger-type simulators. We then implement a compiler pass to exploit the novel functionalities by introducing special instructions governing data movement as part of the quantum circuit. Those instructions target unique capabilities of simulators and have no analogue in actual quantum devices.
To quantify the advantage, we compare the time required to simulate random circuits with and without our optimization. The simulation time is typically halved. 
\end{abstract}



\section{Introduction}
\label{sec:introduction}

The field of quantum computation has recently graduated from scientific research to technology development.
One of the most visible changes has been the creation of increasingly sophisticated software frameworks to enable the execution of algorithms on actual quantum devices \cite{Green2013, Wecker2014, Steiger2018, Qiskit, Cirq, smith2016practical, Killoran2019, Orquestra, Braket}. A major component of the frameworks is the compiler, sometimes also called transpiler, mapper or scheduler depending on its specific role \cite{Javadiabhari2014, Cowtan2019, Childs2019_circuit, Almudever2020}. The compiler is required to translate a relatively abstract algorithm into instructions executable by the electronic controller of the quantum device. Every advanced framework also includes the option of running the algorithm with a simulator, i.e. classical software specialized in simulating quantum circuits.

Simulators have at least two important applications. On one hand they allow to benchmark quantum algorithms without the limitations of existing hardware, in particular finite coherence time \cite{Smelyanskiy2016, Pednault2017, DeRaedt2019_48qub, Haner2017, Khammassi2017, Jones2019, HuaweiHiQ, Villalonga2020}. On the other hand, when equipped with realistic noise models, simulators help characterizing hardware noise and decoherence and, possibly, inform on its origin \cite{OBrien2017, Villalonga2019}.
In spite of their importance, no specialized optimization is integrated in the popular compiler frameworks that mostly deal with restrictions on the qubit connectivity and on the type of available gates. Noticeably, ProjectQ compiler \cite{Steiger2018} accepts multi-qubit operations in the algorithm's description without forcing their decomposition in 1- and 2-qubit gates, a feature suitable for simulation since multi-qubit gates are not native in most quantum hardware.

In this work we present a new compiler pass dedicated to reducing the simulation time for arbitrary quantum circuits. It takes advantage of the specific implementation of the simulator and is, therefore, a backend-aware compilation pass in contrast to hardware-agnostic passes earlier in the compiler chain \cite{Haener2018, Cowtan2019}. We adopt the Intel Quantum Simulator (IQS) \cite{Smelyanskiy2016, Guerreschi2020}, a high-performance simulator exhibiting both multi-threading and distributed parallelization, and extend it by introducing a flexible way to represent the quantum state as a distributed vector. The functionality is available to all users of IQS via its open-source distribution at \url{https://github.com/iqusoft/intel-qs}. Then we create a compiler pass minimizing the communication overhead incurred when multiple computing nodes are involved in the simulation. The optimization is based on the distinction between gates that require inter-node communication and those that do not, minimizing the number of the former ones by means of a greedy search.

While inspired by the implementation of distributed state-vector simulators, similar optimization strategies may be helpful in other scenarios when qubits are divided in two or more categories. For example, one can think of a network of quantum computing chips all involved in the same algorithm. It is easy to envision that the qubits used to communicate between chips differ from in-chip qubits, and therefore the distinction of local/global qubits used in the optimization may be adapted to correspond to qubits for computing/communicating. The scenario of a network of quantum chips is both part of companies road-maps \cite{ionq2020_roadmap} and the starting point of recent research \cite{Rodrigo2021, Haener2021_qmpi} and open-source software \cite{Diadamo2021_qunetsim, Dahlberg2019_simulaqron}.

Previous works on distributed simulators have proposed similar methods to reduce the communication overhead \cite{DeRaedt2007, Haner2017, HuaweiHiQ}. Other works suggested several optimizations based on circuit manipulation, like the notion of fusing several consecutive gates acting on a common subset of qubits \cite{Suzuki2020_qulacs, Qiskit} and then re-express them with fewer operations. Notice that an efficient optimizing pass for gate fusion needs a way to express multi-qubit gates as part of the intermediate representation of the circuit \cite{Haner2017}. In addition, ad-hoc circuit manipulation has been used to simulate the kind of random circuits proposed by Google for quantum supremacy \cite{Arute2019} with a hybrid Schr\"odinger-Feynman simulator developed by NASA \cite{Villalonga2019}, a tensor network-based simulator by Alibaba \cite{Huang2020}, or by leveraging secondary storage as analyzed by IBM \cite{Pednault2019}. However, a central part of our view is that the optimization methods should not be part of a stand-alone simulator, but should occur inside the framework that compiles quantum circuits for different backends. When we describe co-development between compilers and simulators, we have this shift of perspective in mind \cite{Luo2020_yao, Suzuki2020_qulacs}.

To evaluate the efficacy of the specific optimization discussed in this work, we compare the time to simulate random circuits with and without the proposed compiler pass. We vary the ratio of 1- and 2-qubit gates in the circuits and report a reduction by one order of magnitude in the number of gates requiring inter-node communication. This translates in the overall simulation time being typically halved.

\section{Extending Intel Quantum Simulator}
\label{sec:iqs}

IQS is a state-of-the-art Schr\"odinger simulator which represents the full state of $n$ qubits by storing all its $2^n$ complex amplitudes. All operations required by quantum circuits, namely state preparation and several 1- and 2-qubit gates, are implemented as linear manipulation of the amplitudes. Since the state vector grows exponentially in size with the number of qubits, it soon exceeds the memory available in a single computing node. Hence the necessity of distributing the state across multiple nodes, each with its own local memory.

The implementation is presented in references \cite{Smelyanskiy2016, Guerreschi2020}, together with the communication pattern of the Message Passing Interface (MPI) protocol. Here, we provide a high-level summary of the properties exploited by the compiler pass. State preparation and measurement are done with little or no communication and efficiently parallelized over computing nodes and processor's cores (via OpenMP). Gate operations are limited to those gates that update amplitudes in pairs: arbitrary 1-qubit gates and 2-qubit gates in the form of controlled 1-qubit gates. Notice that the 2-qubit gate known as CNOT is part of this set and, thus, IQS can simulate all circuits (arbitrary 1-qubit gates and CNOT allow universal quantum computation).

\begin{figure}[!t]
	\centering
	\includegraphics[width=0.39\textwidth]{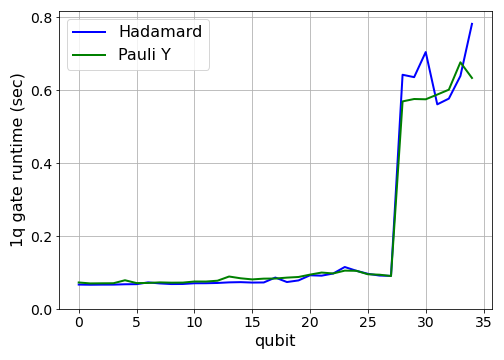} \vspace{3mm} \\
	\includegraphics[width=0.39\textwidth]{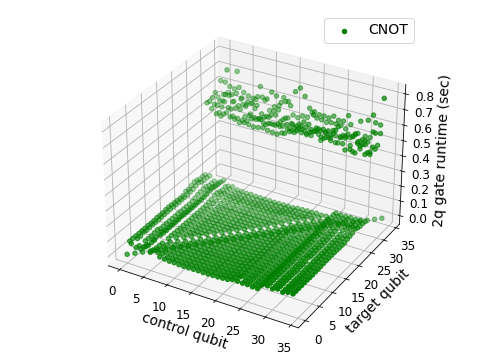}
	\caption{Simulations of $n=35$ qubits using 128 MPI processes, each running on one node of Frontera supercomputer. Each process uses 24 OpenMP threads. \textbf{(Top)} Time to execute a 1-qubit gate as a function of the qubit involved. When the gate is executed on qubit with index $m=n-\lfloor \log_2 k \rfloor$ (with $k$ being the number of MPI processes), the communication between the MPI tasks is happening between sockets of the same node. For qubits with index larger than $m$, communication is between distinct nodes. \textbf{(Bottom)} Time to execute a 2-qubit gate as a function of the involved qubits. CNOT is the controlled Pauli X gate.}
	\label{fig:iqs-benchmarks}
\end{figure}

We divide the qubits into two distinct groups of local, respectively global, qubits \cite{Haner2017}. The number of local \textit{vs} global qubits depends on the number $k$ of MPI processes, specifically there are $\lfloor \log_2 k \rfloor$ global qubits and $m=n-\lfloor \log_2 k \rfloor$ local qubits. Inter-process communication is required by all gates acting on any of the global qubits (apart from the special case of a controlled 2-qubit gate with global control and local target), and the communication overhead is about an order of magnitude. Apart from the communication overhead, the cost of 1- and 2-qubit gates is not significantly different. Figure~\ref{fig:iqs-benchmarks} shows the time required to simulate 1- and 2-qubit gates as a function of the involved qubits. In the numerical experiments, the first $m=28$ qubits are local and is evident that the communication overhead starts from gates on qubits with index 28 or greater. The simulations were run on Frontera supercomputer at the Texas Advanced Computing Center in which each computing node is equipped with 2 sockets of Intel\textsuperscript{\textregistered} Xeon\textsuperscript{\textregistered} Processor 8280 CPU (24 cores per socket).

In the original IQS implementation it is not possible to choose which qubits are local and which are global, by default the first $m$ qubits are the local ones. Consider an arbitrary $n$-qubit state, it can be written in the computational basis as:
\begin{align*}
\ket{\psi} = \sum_{i \in \{0,1\}^n} \alpha_{i} \ket{i_0} \ket{i_1} \dots \ket{i_{n-1}}
\end{align*}
where $i=[i_{n-1} \dots i_1 i_0]$ can be seen as the binary representation of integer $i=\sum_{q=0}^{n-1} i_q 2^q$. IQS stores the quantum state as a distributed vector whose $i$-th component corresponds to amplitude $\alpha_i$. To provide opportunities for fast simulation, we want to increase the flexibility of this representation and, specifically, we allow to change the order in which the amplitudes are stored. Consider an arbitrary permutation $\sigma$ of the qubit indices: under such permutation, qubit $q$ will be associated with bit $\sigma(q)$ of the index of the vector's component. Therefore, amplitude $\alpha_i$ is stored as component $j(i, \sigma)=\sum_{q=0}^{n-1} i_q 2^{\sigma(q)}$ of the state vector.

We introduce the new function \texttt{PermuteQubits}$(\sigma')$ to update the representation of quantum states in IQS. Consider that the current qubit permutation is $\sigma$ and that we want to update it to $\sigma'$. The qubits are reordered in three steps: 1) local qubits (according to $\sigma$) are reordered among themselves, 2) global qubits (according to $\sigma$) are reordered among themselves, and 3) pairs formed by one local and one global qubits (according to either $\sigma$ or, with opposite roles, $\sigma'$) are exchanged. The cost of the three steps is analogous to that of a single local gate, a single global gate, and one global gate per exchanged pair respectively. We discuss the implementation of \texttt{PermuteQubits} in Appendix~A
, together with the pseudo-code to divide every transformation from qubit permutation $\sigma$ to $\sigma'$ in the three steps. As part of this contribution, we added the distributed implementation of SWAP gates to IQS.

The property of local \textit{vs.} global qubits is still valid, but the distinction now depends on permutation $\sigma$. The local qubits are those with index $q$ such that $\sigma(q)<m$. There still are $m=n-\lfloor \log_2 k \rfloor$ local qubits and $\lfloor \log_2 k \rfloor$ global qubits.

\section{Compiler pass to optimize circuit simulation}
\label{sec:pass}

From the benchmarks reported in Figure~\ref{fig:iqs-benchmarks}, it is clear that gates acting on global qubits take much longer to simulate than gates on local qubits alone. We propose a compiler pass that goes through an arbitrary quantum circuit and eliminates the communication overhead from most gates by adding qubit-reordering instructions. Communication is required to execute such instructions in IQS but, when they are chosen carefully, the overhead is reduced.
As we quantify in the next section, even for unstructured circuits the simulation time is halved.

\begin{algorithm}[!b]
	\caption{Compiler pass specialized for IQS}
	\label{alg:compiler-pass}
	\begin{algorithmic}[1]
		\Require The quantum circuit is provided as a DAG describing the logical dependency of the gates. The initial qubit order corresponds to the identity permutation: $\forall q, \; \sigma(q)=q$.
		\Ensure The compiled circuit $C$ is represented by a sequential list of instructions (either gates or calls to \texttt{PermuteQubits}).
		\State $G \gets$ DAG of circuit
		\State $\sigma \gets $identity permutation
		\State $C \gets \{\}$
		\While{$G$.vertices$\neq \emptyset$}
		\For{$v$ in $G$.vertices}
		\If{($v$ has no incoming edges) $\land$ ($v$ acts on local qubits according to $\sigma$)}
		\State \textbf{add} \texttt{ApplyGate}($v$) to $C$
		\State \textbf{remove} $v$ from $G$
		\EndIf
		\EndFor
		\State $\tilde\sigma \gets $identity permutation
		\State $\tilde a \gets 0$
		\For{$l$ in local qubits according to $\sigma$}
		\For{$g$ in global qubits according to $\sigma$}
		\State $\sigma' \gets \sigma \circ (l\,g)$
		\State $a' \gets$ count $G$.vertices on local qubits according to $\sigma'$
		\If{$a'>\tilde a$}
		\State $\tilde \sigma \gets \sigma'$
		\State $\tilde a \gets a'$
		\EndIf
		\EndFor
		\EndFor
		\If{$\tilde a>0$}			
		\State $\sigma \gets \tilde \sigma$
		\State \textbf{add} \texttt{PermuteQubits}($\sigma$) to $C$
		\Else
		\State $v \gets$ vertex of $G$ without incoming edges
		\State \textbf{add} $v$ to $C$
		\State \textbf{remove} $v$ from $G$
		\EndIf
		\EndWhile
	\end{algorithmic}
\end{algorithm}

The compiler pass is described as Algorithm~\ref{alg:compiler-pass}. It assumes that the quantum circuit is provided in terms of a directed acyclic graph (DAG) in which each vertex corresponds to a quantum gate and each edge to a logical dependency: the target vertex (i.e. gate) of the edge must be performed after the source vertex. The DAG may or may not be created considering that quantum gates may commute, but in our case we include commutativity to eliminate unnecessary edges.
While common \cite{Guerreschi2018, Lao2022_timing, Itoko2020}, this is not the only way to describe circuits as DAGs \cite{Sivarajah2020a}. In describing the compiler pass, we will use the terms gate and vertex interchangeably.

The algorithm starts by scheduling all gates that act on local qubits only and do not have unresolved logical dependencies (lines 5:10). The latter condition  corresponds to vertices without incoming edges. When it is not possible to schedule any further gate, the algorithm considers all possible qubit permutations obtained by exchanging one local and one global qubit according to the current qubit permutation $\sigma$ (lines 11:22). There are $m (n-m)\leq n^2/4$ permutations to consider. Each permutation is evaluated depending on the number of gates that it allows to simulate on local qubits, and the permutation with the highest count is recorded as $\tilde \sigma$ in line 18. The compiler add instruction \texttt{PermuteQubit}$(\tilde \sigma)$ if it removes communication from at least one gate, otherwise schedule the first gate without incoming edges even if it requires communication (lines 23:30). The algorithm iterates until all gates are scheduled.

We observe that the compiler pass uses a simplified interpretation of the simulation times reported in Figure~\ref{fig:iqs-benchmarks}. In particular, it considers only whether a qubit is local or global and neglects the fine-grained pattern of the actual simulation time. In addition, it does not consider the actual value of the communication overhead, but is only aware that gates on global qubits takes much longer than gates on local qubits. For example, the plot in Figure~\ref{fig:iqs-benchmarks}(top) would be abstracted as a step function: low simulation time until qubit $m-1$ and high simulation time from qubit $m$ onward. By updating the qubit permutation $\sigma$, the compiler is able to remove certain communication overhead and, therefore, minimize the overall simulation time.
More advanced compiler passes may use the full information from simulator's benchmarks. 
In Appendix~B
, we discuss how to implement the proposed pass in existing compiler pipelines.

\section{Numerical results}
\label{sec:results}

To quantify the advantage provided by the compiler pass, we estimate the time to simulate random circuits before and after their optimization. We consider random circuits of the following form: fix the number of gates, every gate has a probability $p$ to be a 2-qubit gate otherwise with probability $1-p$ it is a 1-qubit gate. Due to stochasticity, the number of two-qubit gates is not strictly determined. The specific type of gate is selected uniformly at random in the sets $\{H, Y\}$ and $\{CNOT\}$ for 1- and 2-qubit gates respectively. Finally the qubits involved in each gate are chosen uniformly at random. Recall that the simulation time is largely independent on the specific type of gate and the choice is therefore arbitrary and inconsequential.

\begin{figure}[!t]
	\centering
	\hspace{0.5mm}
	\includegraphics[width=0.43\textwidth]{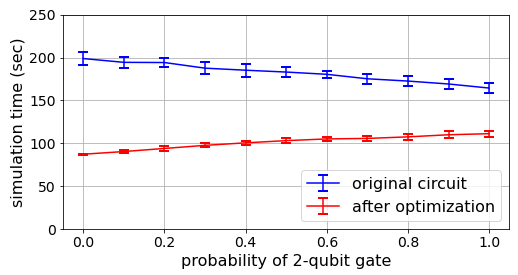} \vspace{3mm} \\
	\includegraphics[width=0.425\textwidth]{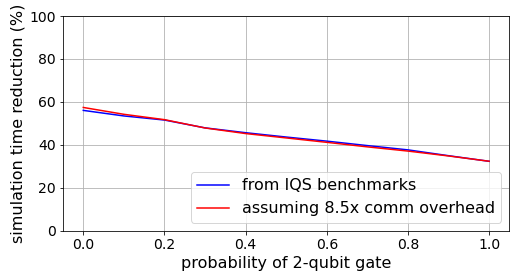}
	\caption{Effect of the compiler pass to optimize quantum circuits for simulation with IQS. \textbf{(Top)} Simulation time of random circuits of 1050 gates on $n=35$ qubits. The timings have been estimated for the same configuration used in Figure~\ref{fig:iqs-benchmarks}: 128 computing nodes of Frontera supercomputer at TACC. There are $m=28$ local qubits. Datapoints represent the average over 20 random circuits and the vertical bars indicate one standard deviation. \textbf{(Bottom)} Reduction of the simulation time between original and optimized circuit. The blue line is based on direct benchmarks of IQS, the red line is obtained by assuming that operations requiring communication have an overhead of $8.5\times$ with respect to operations implementable locally. This overhead is in line with the benchmarks in Figure~\ref{fig:iqs-benchmarks}.}
	\label{fig:loop-on-prob}
\end{figure}

The random circuits we consider are simply a sequence of gates chosen at random, and do not aim at reproducing a Haar-random unitary on $n$ qubits. They have been chosen to reflect the common situation in which programmers of quantum algorithms are not familiar with the design of the simulator and, thus, do not tailor their code (via gate selection or qubit indexing) for a specific simulator. In addition, while random circuits are not associated with a specific application, they are representative of algorithms like the Quantum Approximate Optimization Algorithm to solve combinatorial problems on class of random instances \cite{Farhi2014} or the dynamical evolution of spin systems with random interactions \cite{Salathe2015}. In Appendix~C
, we present results for circuits having a clear pattern of 2-qubit gates and show evidence supporting an even greater reduction in simulation time.

In the first study, we fix the number of qubits to $n=35$ and vary the probability $p$. Despite only a fraction $7/35=1/5$ of the qubits being global, the simulation time is reduced by between 32.3\% (for $p=1$) and 56.0\% (for $p=0$), according to the formula $1-t_\text{opt}/t_\text{orig}$ in which $t_\text{orig}$ is the simulation time for the original circuit and $t_\text{opt}$ that for the optimized circuit. The results are shown in Figure~\ref{fig:loop-on-prob} and have been obtained by averaging over 20 random circuits of $(30\,n)$ gates on $n=35$ qubits. The diminished advantage when $p$ increases is not related to the different probability that a 2-qubit gate requires communication compared to a 1-qubit gate, in fact for CNOT gates communication is required if and only if the target is global irrespective of the control qubit. We attribute the diminished reduction to the fact that a 2-qubit gate (requiring communication) excludes a larger part of the remaining circuit from execution since it has, on average, more outgoing edges in the DAG description of the circuit (see condition in line 6 of Algorithm~\ref{alg:compiler-pass}).

The simulation time has been estimated by summing up the time required to simulate every gate separately.
We start by benchmarking the execution of each type of gate acting on every qubit or pair of qubits. We also benchmark the time to reorganize the classical data representing the quantum state following a call to \texttt{PermuteQubits}. This can be achieved by benchmarking the time to execute a single SWAP gate on every pair of qubits since each new permutation is related to the previous one by the composition with a single 2-cycle or transposition (see line 15 in Algorithm~\ref{alg:compiler-pass}).
Notice that, unlike from the circuit optimization that is based on the dicotomy local \textit{vs.} global qubits, to estimate the simulation time of both the original and optimized circuits we consider the full information content provided by benchmarks like those in Figure.~\ref{fig:iqs-benchmarks}.

\begin{figure}[!t]
  \centering
  \includegraphics[width=0.43\textwidth]{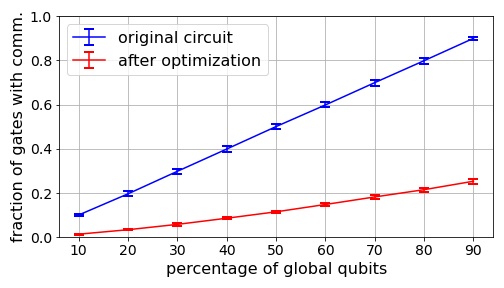} \vspace{3mm} \\
  \includegraphics[width=0.43\textwidth]{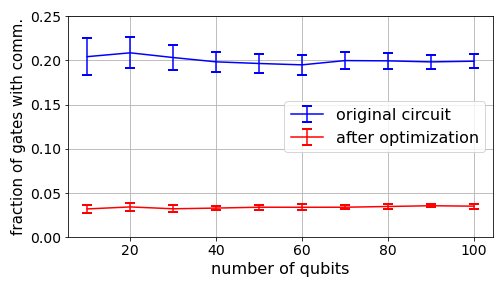}
  \caption{Effect of the compiler pass to optimize quantum circuits for simulation with IQS. Fraction of gates requiring communication for the original (blue) and optimized (red) circuit. \textbf{(Top)} Fixing $n=50$, we vary the percentage of global qubits from 10\% to 90\%. \textbf{(Bottom)} Fixing the ratio global \textit{vs.} total qubits to 1:5, we vary $n$. In both panels, datapoints are averaged over 20 random circuits and the vertical bars indicate one standard deviation. The random circuits have $(30\, n)$ gates.}
  \label{fig:loop-on-numq}
\end{figure}

In the second study, we fix the probability of 2-qubit gates to $p=0.3$ and vary the number of local and global qubits. This choice corresponds to about two 1-qubit gates every 2-qubit operation. In Figure~\ref{fig:loop-on-numq} we report the number of gates causing communication overhead (here including also the qubit reordering operations) as a fraction of the total number of original operations. In the top panel, the horizontal axis represents the percentage of qubits that are global when the total number of qubits is $n=50$. This simulation size is at the threshold of simulability, even considering using all computing nodes of current supercomputers. However, we show in the bottom panel that, apart from small-size effects for $n \leq 30$, the behavior is independent on $n$. This allows us to estimate the advantages of the optimization at extremely large scale (even beyond the simulability threshold expected around 50 qubits). Specifically, when $1/5$ of the qubits are global, the fraction of gates requiring communication is reduced from $\sim 20\%$ to $\sim 3.5\%$ independently of the number of qubits. Assuming an overhead of about $8.5\times$ for the communication (see bottom panel of Figure~\ref{fig:loop-on-prob}), this would correspond to a reduction of simulation time by $1-t_\text{opt}/t_\text{orig}=1-(0.965+0.035\times 8.5)/(0.8+0.2\times 8.5)\simeq 49.5\%$. Notice that our simple optimization may be improved, but with limited gain since even the impossible case of completely eliminating communication would lead to a reduction of $1-t_\text{no-comm}/t_\text{orig}=1-1/(0.8+0.2\times 8.5)\simeq 60\%$ in this scenario.

The latter observation raises the question whether a $\sim 2 \times$ speedup in simulation is an important benchmark. We believe so, especially considering that the typical use case of IQS, and of increasingly more simulators due to the progress of quantum hardware technology, are large-scale simulations running on hundreds or thousands of computing nodes for an extended period. Halving the simulation costs, not only the temporal cost but also the monetary and energy cost \cite{Villalonga2020}, with no additional user's overhead is certainly remarkable.


\section{Conclusion and outlook}
\label{sec:conclusion}

Quantum circuits are compiled for execution on actual quantum hardware, but an important feature of quantum computing frameworks is the possibility of simulating algorithms on classical computers. We propose a new compiler pass specialized for circuit simulation. Our results demonstrate that the co-development and integration between circuit compiler and simulator is largely beneficial and reduces the simulation time in half. The integration requires dedicated functions in both software programs: on one hand, we extend Intel Quantum Simulator, a high-performance, distributed and open-source simulator, with flexible data structures and qubit re-ordering functionalities; on the other hand, we propose a novel compiler pass based on the minimization of the number of operation requiring communication between computing nodes. We believe that compiler passes specialized for circuit simulation will become an essential component of every quantum computation framework. 


\section*{Acknowledgements}
The author acknowledges the Texas Advanced Computing Center (TACC) at The University of Texas at Austin for providing HPC resources that have contributed to the research results reported in this work. \url{http://www. tacc.utexas.edu}
The author thanks Fabio Baruffa, Justin Hogaboam, and Nicolas P.D. Sawaya for helpful discussions, and Salvatore Mandr\`a for comments on a previous version of the manuscript.


\bibliographystyle{quantum}
\bibliography{references}

\vspace{5mm}
\bigskip\noindent\makebox[\linewidth]{\resizebox{0.5\linewidth}{1pt}{$\bullet$}}\bigskip
\vspace{8mm}

\appendix

\section{Appendix A:\\Transformation between two qubit permutations}
\label{app:sec:transformation}

Consider the current qubit permutation $\sigma$ such that amplitude $\alpha_i$
of the computational basis state $\ket{i_{n-1}}\dots\ket{i_1}\ket{i_0}$ corresponds
to the component $j=\sum_{q=0}^{n-1} i_q 2^{\sigma(q)}$ of the state vector.
We want to update the qubit permutation to $\sigma'$ taking into account that
$m$ of the $n$ qubits are local and the remaining $n-m$ are global. The IQS method
implementing the update proceeds in three steps: The first one reorders local qubits only,
the second one reorders global qubits only, and the third step exchanges pairs formed
by one local and one global qubits.

The three steps have separate implementations: reordering local qubits is obtained by copying
and rearranging data in each local memory independently, reordering global qubits is
obtained by swapping the local memory of two distinct processes, and exchanging a local
and a global qubit corresponds to the execution of a SWAP gate from the point of view of data
movement. A SWAP-like gate is implemented via the same communicate-compute-communicate approach
of controlled 1-qubit gates described in \cite{Smelyanskiy2016}, but with a different pattern
of inter-process communication.

Below, we provide pseudocode to divide the transformation $\sigma \mapsto \sigma'$
in three transformations that, composed sequentially, correspond to the desired one.
For clarity of terminology, we will say that qubit $q$ is mapped to position $\sigma(q)$ and,
viceversa, that position $p$ is mapped to qubit $\sigma^{-1}(p)$. Inverting the permutation
is efficient and takes $\mathcal{O}(n)$ operations. Algorithm~\ref{alg:permute-qubits} constructs the intermediate
permutations such that $\sigma \mapsto \sigma_1 \mapsto \sigma_2 \mapsto \sigma'$ satisfies
the criteria expressed above.

\begin{algorithm}[h]
	\caption{Decompose transformation $\sigma \mapsto \sigma'$}
	\label{alg:permute-qubits}
	\begin{algorithmic}[1]
		\item [Three steps: $\sigma \mapsto \sigma_1 \mapsto \sigma_2 \mapsto \sigma'$]
		\State $\sigma \gets$ current qubit permutation
		\State $\sigma' \gets$ desired qubit permutation
		\State $\sigma_1 \gets$ identity permutation
		\State $\sigma_2 \gets$ identity permutation
		\For{$p = 0,1,\dots,m-1$}
			\State $p' \gets \sigma'( \sigma^{-1} (p))$
			\While{$p' \geq m$}
				\State $p' \gets \sigma'( \sigma^{-1} (p'))$ 
			\EndWhile
			\State $\sigma_1^{-1}(p') \gets \sigma^{-1}(p)$
		\EndFor
		\For{$p = m, m+1,\dots,n-1$}
			\State $p' \gets \sigma'( \sigma_1^{-1} (p))$
			\While{$p' < m$}
				\State $p' \gets \sigma'( \sigma_1^{-1} (p'))$ 
			\EndWhile
			\State $\sigma_2^{-1}(p') \gets \sigma_1^{-1}(p)$
		\EndFor
	\end{algorithmic}
\end{algorithm}

\section{Appendix B:\\Implement optimization pass in existing compiler pipelines}
\label{app:sec:implementation}

In Algorithm~\ref{alg:compiler-pass} of the main text, we provide the pseudocode for the optimization pass suggested in this work. While we do not provide an explicit implementation, we would like to comment on how it could be integrated in existing compiler pipelines with minimal modifications.

From the pseudocode it should be clear that the effect of ``\textbf{add} \texttt{PermuteQubits}$(\sigma)$ to C'' in line 25 is to change the qubit order by swapping only two qubits, one local and the other global. Therefore, one can introduce two separate instructions in the compiler pipeline: SWAP and Emulated-SWAP. While SWAP corresponds to the usual 2-qubit gate (this instruction is typically present in any quantum compiler), Emulated-SWAP can be seen as the same operation executed in two different ways: For hardware backends it corresponds to a usual SWAP, for Intel Quantum Simulator backend it updates the qubit order without data movement.

At this point, ``\textbf{add} \texttt{PermuteQubits}$(\sigma)$ to C'' corresponds to adding a SWAP instruction followed by an Emulated-SWAP on the same pair of qubits. The intermediate representation of the quantum circuit is easy to interpret since Emulated-SWAP can be treated as any other 2-qubit gate and, if needed, the code can also run for hardware backends or other simulators.

\section{Appendix C:\\Reduction of the simulation time for QFT-like circuits}
\label{app:sec:qft_like}

In the main text we applied the proposed compiler pass to random circuits. Being part of the compiler pipeline, the pass can be applied to any quantum circuit without additional effort by the user writing the quantum algorithm. However, its effectiveness depends on the pattern of 2-qubit gates. We expect a large reduction in simulation time for circuits which can be divided in a small number of subcircuits, each involving only a fraction of the total qubits. This would allow to permute the qubits not involved in the subcircuit at most once during the subcircuit, making them global qubits if they were local and avoiding any further permutation involving them during the subcircuit.

For our study in Section \textbf{\textit{Numerical results}}, we chose circuits with 2-qubit gates involving qubits chosen uniformly at random. In this case, it is difficult to identify the subcircuit structure discussed above and therefore we believe that the random circuit benchmarks reflect unfavorable situations.

To confirm this intuition, we consider the opposite scenario of circuits presenting a clear pattern of 2-qubit gates. As an example, we analyze QFT-like circuits (QFT stays for Quantum Fourier Transform) in which the controlled-Rz gates are substituted by CNOT gates. The reason behind the substitution is that controlled-Rz gates are diagonal in the computational basis and do not require communication, while CNOT gates do require communication as benchmarked in  Figure~\ref{fig:iqs-benchmarks}. The circuits are of the form illustrated in Figure~\ref{fig:qft-circuit} below. 

\begin{figure}[h!]
	\centering
	\vspace{2mm}
	\includegraphics[width=0.48\textwidth]{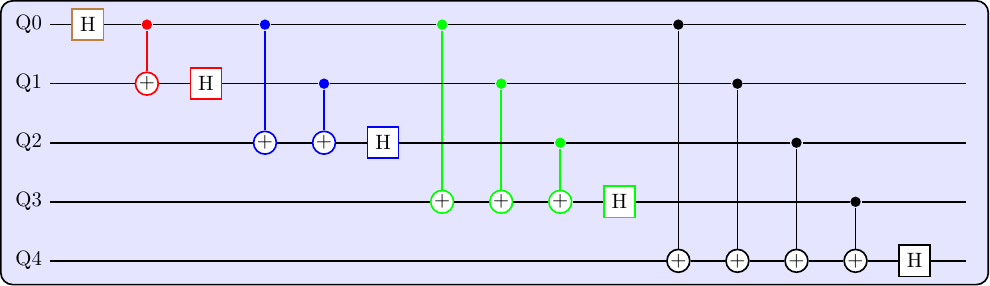}
	\caption{Quantum circuit with the same structure as the Quantum Fourier Transform circuit, but with CNOT gates substituting controlled-Rz rotations.}
	\label{fig:qft-circuit}
\end{figure}

\begin{figure}[h!]
	\centering
	\includegraphics[width=0.44\textwidth]{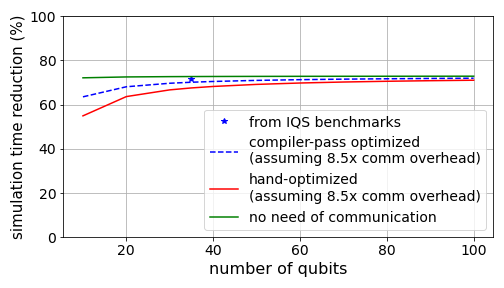}
	\caption{Effect of the compiler pass to optimize QFT-like quantum circuits for simulation with IQS. The blue star uses the benchmark results from Figure~\ref{fig:iqs-benchmarks}, while the dashed blue line uses a simplified model in which operations requiring communication have an overhead of 8.5x with respect to operations implementable locally (see caption of Figure~\ref{fig:loop-on-prob}). The green line represents the limiting case in which no operation would require communication. For all circuit sizes, it has been assumed that $20\%$ of the qubits are global.}
	\label{fig:qft-results}
\end{figure}

With the trivial qubit ordering in which low-index qubits are local and high-index qubits are global, one can compute the number of gates requiring communication as a function of the number of qubits $n$ and the fraction of local qubits $m/n$. Additionally, one can observe that at most $2 (n-m)$ qubit permutations are needed to eliminate all communication overhead (two permutations for each of the same-color gates with Hadamard on a global qubit). Figure~\ref{fig:qft-results} clarifies that the pass we propose is particularly effective for QFT-like circuits.

\end{document}